\documentstyle[12pt]{article}   
\begin{document}    
\baselineskip=24.5pt    
\setcounter{page}{1}         
\topskip 0 cm    
\vspace{1 cm}    
\centerline{\LARGE\bf  A Higher-dimensional Origin of}
\centerline{\LARGE\bf  the Inverted Mass Hierarchy for Neutrinos }
\vskip 1 cm
\centerline{\large M. Tanimoto$^1$ and T. Yanagida$^2$}
\vskip5mm
\centerline{$^1$ Department of Physics, Niigata University, Niigata 950-2181,
Japan}
\centerline{$^2$ Department of Physics, University of Tokyo,  Tokyo 113 0033, 
Japan}

 
\vskip 4 cm
\noindent
{\large\bf Abstract}

We present  successful lepton mass matrices with an inverted mass hierarchy 
for neutrinos,  which follow from 
a geometrical structure of a (1+5) dimensional space-time where two extra
dimensions are compactified on the ${\bf T^2/Z_3}$ orbifold.
A  ${\bf 5}^*$ and a right-handed neutrino $N$  in each family are 
localized on each of the equivalent three fixed points of the orbifold 
while three ${\bf 10}$'s and Higgs doublets $H_u$ and $H_d$ live in the bulk. 
An $S_3$ family symmetry is assumed on three ${\bf 5}^*$'s and on three $N$'s,
since the three fixed points are  equivalent to one another.
The Higgs field $\phi$ responsible for the B-L breaking is localized on one of 
the three fixed points,
which  generates the inverted hierarchy for the neutrino masses. 
The  baryon asymmetry is well explained in the non-thermal leptogenesis 
via inflaton decay. We emphasize that the present model predicts the effective
 neutrino mass, $\langle m\rangle_{ee}$, responsible for 
neutrinoless double beta decays as $\langle m\rangle_{ee}\simeq 50$ meV.
This will be accessible to future experiments.

\newpage

\section{Introduction}

It is a challenging and important task to find an origin of the observed 
hierarchies in mass matrices for quarks and leptons, since it may provide 
us with a clue to a more fundamental theory beyond the standard model.
The most remarkable observation is that the weak mixing angles in the 
lepton sector are of the order 1 while those in the quark sector are 
small. This indicates that nature of left-handed lepton doublets is very
different from that of left-handed quark doublets, which means that
there is a big disparity between multiplets ${\bf 5}^*$'s and ${\bf 10}$'s 
in the SU(5) grand unification model.

It has been pointed out \cite{Geometry} that a (1+5) dimensional space-time 
may explain the above disparity if extra two dimensions are compactified 
on the ${\bf T^2/Z_3}$ orbifold. This orbifold has three equivalent fixed
points. Here, a ${\bf 5}^*$ and a right-handed neutrino $N$ in each 
family are localized on each of the  three fixed points 
of the orbifold, 
while three ${\bf 10}$'s live in  the bulk. One may obtain naturally  
successful democratic mass matrices for quarks and leptons, provided that 
the three ${\bf 10}$'s, Higgs doublets and a Higgs $\phi$ responsible for the 
B-L symmetry breaking are distributed homogeneously in the bulk. Thus, this 
theory may provide a model for the almost degenerate neutrino masses.

It has been, furthermore, shown \cite{IIB} that an extension of the model
 to the type 
IIB string theory may explain not only the above geometrical structure of 
the wave functions for ${\bf 5}^*$'s, $N$'s and ${\bf 10}$'s but also 
the presence of Higgs doublets $H_u$ and $H_d$ in the bulk. 
However, the nature of the B-L breaking is 
not clearly understood. It is remarkable that if the Higgs $\phi$ for the B-L
breaking is localized on one of the three fixed points, one of $N$ acquires 
a superheavy Majorana mass generating an inverted hierarchy 
in the neutrino mass spectrum. 
The purpose of the present paper is to discuss this possibility.

Our analysis shows that the present model is consistent with all 
observations on neutrino masses and mixing angles. Further we show that 
the matrix element 
responsible for neutrinoless double beta decays is predicted 
as $\langle m\rangle_{ee}\simeq 50$ meV. 
The  prediction will be tested in future  experiments.

 In section 2, we discuss the neutrino and charged lepton mass matrices,
where the inverted neutrino mass spectrum is naturally obtained.
In  section 3, the non-thermal leptogenesis via inflaton decay is discussed 
in the present model. The summary is devoted to section 4.
 
\section{Neutrino Masses and  Mixings}

Our discussion is based on the model proposed in \cite{Geometry, IIB}.
Here, a ${\bf 5}^*$ and a right-handed neutrino $N$  in each family are 
localized 
on one of the equivalent three fixed points of the ${\bf T^2/Z_3}$ orbifold 
while three ${\bf 10}$'s and Higgs doublets $H_u$ and $H_d$ live in the bulk. 
We assume in this paper that the $\phi$ for the B-L breaking is localized on 
the fixed point on which the ${\bf 5}^*$ and $N$ in the third family reside. 
Thus, the $N_3$ in the third family has a very large Majorana mass compared
 with those of other $N$'s.

We consider an interchange symmetry between the fields on the orbifold fixed 
points,  since the three fixed points are  equivalent to one another. 
That is, we assume an
$S_3$ family symmetry acting on three ${\bf 5}^*$'s and on three $N$'s.
Then, the neutrino and charged lepton mass matrices are given in terms of 
a few parameters.

However, we have to introduce breakings of the $S_3$ symmetry to obtain 
realistic mass matrices. We assume two sources of the breaking. One is a 
localization of the wave function of the $\phi$ and 
the other is small distortions
of the wave functions of the three ${\bf 10}$'s and the doublet Higgses from 
homogeneous forms in the bulk.

\subsection{Neutrino  Mass Matrix}

Let us begin by discussing  the Dirac neutrino mass matrix,
which is given by  the Yukawa coupling matrix  of $N{\bf 5}^* H_u$.
There are two independent matrices which are invariant of the $S_3$ symmetry \cite{FTY,FX}.
Namely, the $S_3$ invariant Dirac mass matrix is given by \cite{FTY}
\begin{equation}
h_\nu=h_0\left [\left (\matrix{1& 0& 0\cr 0 &1 &0\cr 0 & 0 &1 \cr      } \right )
+\left ( \matrix{0& \epsilon & \epsilon \cr
                  \epsilon  & 0 & \epsilon\cr
                  \epsilon  & \epsilon  & 0 \cr
                                         } \right ) \right ].
\end{equation}
\noindent
Notice that the parameter $\epsilon$ is suppressed by separation of
the three fixed points. 
We expect $\epsilon \simeq {\rm exp}(-\ell_*\times M_*)$
where $\ell_*$ and $M_*$ are the distance between different fixed points and
a mass scale of the fundamental theory, respectively \cite{Ark}. For instance, we have 
$\epsilon\simeq  0.001$ for $\ell_*\times M_*\simeq 7$.

We have assumed, so far, a homogeneous distribution of the Higgs fields $H_u$
in the bulk. A small distortion of the Higgs wave function may induce a violation 
of the $S_3$ symmetry. The breaking effects appear first in the diagonal 
elements of the above mass matrix and the effects in the $\epsilon$ term may 
be negligible for our discussion.
Therefore, the neutrino Dirac mass matrix is given by 
\begin{equation}
h_\nu=h_0\left [\left (\matrix{1& 0& 0\cr 0 &1+\delta_1 &0\cr 0 & 0 &1+\delta_2 \cr      } \right )
+\left ( \matrix{0& \epsilon & \epsilon \cr
                  \epsilon  & 0 & \epsilon\cr
                  \epsilon  & \epsilon  & 0 \cr
                                         } \right ) \right ] 
=h_0\left ( \matrix{1& \epsilon & \epsilon \cr
                  \epsilon  & 1+\delta_1 & \epsilon\cr
                  \epsilon  & \epsilon  & 1+\delta_2 \cr
                                         } \right )\ .
\end{equation}
\noindent
Here, we take a basis where all the diagonal elements are real 
while the off diagonal elements $\epsilon$ are  complex.

The Majorana mass matrix $M_R$ for the right-handed neutrino $N_i ~(i=1-3)$ 
is determined also
by localization properties of $N_i$ and $\phi$ fields.
Since the $\phi$  is assumed to reside on one of the three fixed points where the third family
$N_3$ is localized, the (3,3) element of the Majorana mass matrix dominates 
over other elements. 
Then, the Majorana mass matrix is given in terms of three independent 
suppression factors $\epsilon_1$,  $\epsilon_2$ and $\epsilon_3$ as follows:
\begin{equation}
M_R= M_0\left ( \matrix{\epsilon_2& \epsilon_3^2 & \epsilon_1 \cr
                  \epsilon_3^2  & \epsilon_2 & \epsilon_1\cr
                  \epsilon_1  & \epsilon_1  & 1\cr
                                         } \right )\ ,
\label{rightmatrix}
\end{equation}
\noindent
where we always take the three diagonal elements to be real.
In the following discussion we assume all elements in the matrix  $M_R$ to be real, for simplicity. 
Notice that the suppression factors come from the separation of the distinct fixed
points and hence the parameters $\epsilon_i~(i=1,2,3)$ are the same order of the magnitude 
of $|\epsilon |$.

 Through the  seesaw mechanism \cite{Seesaw, Seesaw2},
the  effective neutrino mass matrix is given, in the leading order
of $\epsilon,\ \epsilon_i$ and $\delta_i$, by
\begin{equation}
 M_\nu= h_\nu^T M_R^{-1}h_\nu =
\frac{h_0^2}{\epsilon_2 M_0}\left (\matrix{1-\frac{\epsilon_1^2}{\epsilon_2}& 
  2\epsilon+\tilde \epsilon & \epsilon-\epsilon_1 \cr
2\epsilon+\tilde \epsilon& 1+2\delta_1-\frac{\epsilon_1^2}{\epsilon_2} 
& \epsilon-\epsilon_1\cr
  \epsilon-\epsilon_1  & \epsilon-\epsilon_1 & \epsilon_2 \cr
                                         } \right )\ ,
\label{mass}
\end{equation}
\noindent 
where 
\footnote{A general analysis  including other possible phases
is prepared for  CP violating phenomena at low energies.}
\begin{equation}
\epsilon=|\epsilon| e^{i \varphi} \ , \qquad
\tilde \epsilon =\frac{\epsilon_1^2 - \epsilon_3^2}{\epsilon_2} \ .
\end{equation}
The neutrino mass eigenvalues are obtained  as
\begin{equation}
m^2_{1}\simeq   \frac{h_0^2}{\epsilon_2^2 M_0^2} \ , \qquad
m^2_{2}\simeq   \frac{h_0^2}{\epsilon_2^2 M_0^2} 
             (1+4 \delta_1-4 \frac{\epsilon_1^2}{\epsilon_2})\ , \qquad
m_3^2\simeq  \frac{h_0^2}{\epsilon_2^2 M_0^2} \epsilon_2^2 \ ,
\label{mass2}
\end{equation}
\noindent
which are the spectrum called as  inverted mass hierarchy.
Therefore, the ratio of the solar neutrino mass scale and 
the atmospheric neutrino mass scale is  given in terms
of $\delta_1$ and $\epsilon_i$,
\begin{equation}
\frac{\Delta m^2_{\rm solar}}{\Delta m^2_{\rm atm}}=
\frac{m_2^2-m_1^2}{m_1^2-m_3^2}
\simeq 4(\delta_1  - \frac{\epsilon_3^2}{\epsilon_2}) \ .
\end{equation}
The experimental data of $\Delta m^2_{\rm solar}\simeq 8\times
10^{-5}{\rm eV^2}$ and  $\Delta m^2_{\rm atm}\simeq 2.2\times
10^{-3}{\rm eV^2}$ \cite{Solar,Kamland,Atmospheric} gives us a constraint,
\begin{equation}
\delta_1\simeq  9\times 10^{-3} \ ,
\label{delta}
\end{equation}
where $\epsilon_i$ are neglected against  $\delta_1$.
The condition  $\epsilon_i\ll \delta_1$ is required to get the observed
 $\sin\theta_{12}$ in eq. (\ref{mix}) of the subsection  {\bf 2.3}.

The unitary matrix, which diagonalizes the neutrino mass matrix of  
eq.(\ref{mass}) ( $V_\nu^T  M_\nu V_\nu = M_{\rm diagonal}$ ), 
is given approximately by
\begin{eqnarray}
&& V_\nu \simeq   \left(
 \matrix{ 1 & \frac{2|\epsilon|\cos\varphi+\tilde \epsilon}{2\delta_1}
 &  {\mathcal O}(\epsilon,\epsilon_i)\cr
    -\frac{2|\epsilon|\cos\varphi+\tilde \epsilon}{2\delta_1} & 1& {\mathcal O}(\epsilon,\epsilon_i)\cr
 {\mathcal O}(\epsilon,\epsilon_i)&   {\mathcal O}(\epsilon,\epsilon_i)& 1 \cr
                                         } \right )  \ .
\label{angle}
\end{eqnarray}
\noindent
It should be noticed that
the CP violating   phase $\varphi$ appears here 
at the order of $\epsilon$.
The Majorana phases are also estimated  from the mass matrix of 
eq.(\ref{mass}), 
where the diagonal elements are real and off diagonal ones have imaginary
parts of the order of  $\epsilon$ in the unit of $h_0^2/(\epsilon_2 M_0)$. 
After diagonalizing the mass matrix, 
the first and second masses $m_1$ and $m_2$ get phases
of the order of  $\epsilon$ while  $m_3$ has  a phase 
of  the order of  $\epsilon^2$. 
Therefore, the Majorana phases are at most 
of  the order of  $\epsilon$, which are very small.

\subsection{Charged Lepton Sector}

When the three ${\bf 10}$'s and the Higgs multiplet $H_d$ distribute 
homogeneously in the bulk, one obtains the following  democratic mass matrix 
for charged leptons:
\begin{equation}
M_\ell=\frac{m_0}{a+b+c}
\left (\matrix{a& a& a \cr b& b& b\cr  c & c & c \cr} \right )  \ .
\end{equation}
\noindent
In this mass matrix, the left-handed mixing is independent of
 values $a$, $b$ and $c$ while the right-handed one  depends
on these values.
We take $a=b=c=1$, for simplicity, since our concern
is the left-handed mixing angles.

Now we introduce distortions of  wave functions of the ${\bf 10}$'s 
in the bulk. We assume that dominant effects appear on the diagonal elements
of the above matrix. Then, the charged lepton mass matrix is given by , 
\begin{equation}
M_\ell=\frac{m_0}{3}
\left (\matrix{1& 1& 1 \cr 1& 1& 1\cr  1 & 1 & 1 \cr} \right )
+ \left (\matrix{\delta_{\ell 1}& 0&0 \cr 0& \delta_{\ell 2}& 0\cr
  0& 0  &\delta_{\ell 3}\cr} \right )  \ ,
\label{demo}
\end{equation}
\noindent where the second term of the right-handed side is the $S_3$ 
breaking  terms coming from the distortions of the ${\bf 10}$ fields. 
We take all $\delta_{\ell i}$ to be real, for simplicity.
This form of the mass matrix is used in ref.\cite{Koide,FTY}. Here, we have 
assumed that the effects of distortion of the Higgs field $H_d$ are 
negligibly small.

The matrix of eq.({\ref{demo}) is diagonalized by $V_\ell=F_0 L_\ell$, where
\begin{eqnarray}
&& F_0= 
\left(\matrix{\frac{1}{\sqrt{2}} & \frac{1}{\sqrt{6}} & \frac{1}{\sqrt{3}}\cr
       -\frac{1}{\sqrt{2}} &  \frac{1}{\sqrt{6}} & \frac{1}{\sqrt{3}} \cr
               0       & -\frac{2}{\sqrt{6}} & \frac{1}{\sqrt{3}} \cr
                                         } \right )  \ , \nonumber \\
&& F_\ell \simeq   \left(
 \matrix{\cos\theta_\ell&\sin\theta_\ell&\lambda_\ell\sin 2\theta_\ell\cr
       -\sin\theta_\ell & \cos\theta_\ell &-\lambda_\ell\cos 2\theta_\ell \cr
 -\lambda_\ell\sin 3\theta_\ell&  \lambda_\ell\cos 3\theta_\ell & 1 \cr
                                         } \right )  \ , 
\end{eqnarray}
\noindent with
\begin{eqnarray}
&&\tan 2\theta_\ell \simeq \sqrt{3}
\frac{\delta_{\ell 2}-\delta_{\ell 1}}
{2\delta_{\ell 3}-\delta_{\ell 2}-\delta_{\ell 1}} \ ,
\qquad\qquad \lambda_\ell\simeq \frac{1}{3\sqrt{2}}\frac{\xi_\ell}{m_0}\ ,
\nonumber\\
&&\xi_\ell=\sqrt{(2\delta_{\ell 3}-\delta_{\ell 2}-\delta_{\ell 1})^2+
3 (\delta_{\ell 2}-\delta_{\ell 1})^2} \ .
\end{eqnarray}
The mass eigenvalues are given by
\begin{eqnarray}
&&m_e=\frac{1}{3}(\delta_{\ell 1}+\delta_{\ell 2}+\delta_{\ell 3})
-\frac{1}{6}\xi_\ell \ , \nonumber\\
&&m_\mu=\frac{1}{3}(\delta_{\ell 1}+\delta_{\ell 2}+\delta_{\ell 3})
+\frac{1}{6}\xi_\ell \ ,\nonumber\\
&&m_\tau=m_0+\frac{1}{3}(\delta_{\ell 1}+\delta_{\ell 2}+\delta_{\ell 3})\ .
\end{eqnarray}
\noindent 
If we take $\delta_{\ell 1}+\delta_{\ell 2}=0$ and 
$\delta_{\ell 2}\ll \delta_{\ell 3}$, for simplicity  \cite{FTY},
we get following mixings  in terms of the charged lepton masses:
\begin{equation}
\sin\theta_\ell \simeq -\sqrt{\frac{m_e}{m_\mu}} \ , \qquad
 \lambda_\ell\simeq \frac{1}{\sqrt{2}} \frac{m_\mu}{m_\tau}  \ ,
\label{mix0}
\end{equation}
\noindent which is used in our numerical calculations.
It may be important to note that the condition
 $\delta_{\ell 1}+\delta_{\ell 2}=0$
 is crucial for the prediction of  $\sin\theta_\ell$.
For example,  if we take another condition
 $\delta_{\ell 1}\ll \delta_{\ell 2}\ll  \delta_{\ell 3}$,
we have $\sin\theta_\ell \simeq -m_e/\sqrt{3}m_\mu$,
which is much smaller than the one in eq.(\ref{mix0}).
Therefore,  the prediction of $\sin\theta_{13}$ strongly
depends on the assumption of $\delta_{\ell i}$ as seen in eq.{(\ref{mix})}.

\subsection{Lepton  Flavor Mixings }

The lepton flavor  mixing matrix $U$ \cite{MNS}
is obtained as
\begin{eqnarray}
 U = F_\ell^{\dagger} F_0^{\dagger} V_\nu \ ,
\end{eqnarray}
\noindent
which gives
\begin{eqnarray}
 && \sin\theta_{12}\simeq -\frac{1}{\sqrt{2}}+\sqrt{\frac{m_e}{m_\mu}}+
 \frac{1}{\sqrt{2}}\frac{2|\epsilon|\cos\phi+\tilde \epsilon}{2\delta_1} \ ,
\nonumber\\
&& \sin\theta_{13}\simeq \frac{2}{\sqrt{3}}\sqrt{\frac{m_e}{m_\mu}} \ , 
\nonumber\\
&&\sin\theta_{23}
\simeq -\frac{2}{\sqrt{3}}+\frac{1}{\sqrt{6}}\frac{m_\mu}{m_\tau} \ ,
\label{mix}
\end{eqnarray}
\noindent where
 $\theta_{ij}$ correspond to the mixing angles  in the conventional 
parameterization of the mixing matrix in  PDG \cite{PDG}.
For example, we take a  constraint in eq.(\ref{delta}), $\delta_1=0.009$,
 with $|\epsilon|=0.002$ and $|\varphi|=\pi/4$, which leads to
\begin{eqnarray}
  \sin^2 \theta_{12}=0.30\ , \qquad \sin^2 2\theta_{23}=0.93\ , 
\qquad \sin^2\theta_{13}=0.0025 \ .
\end{eqnarray}
\noindent Those are consistent with 
  the result of three flavor analyses
 on the experimental data with   $3\sigma$   in~\cite{tortola,fogli},
\begin{eqnarray}
& & 7.2 \times 10^{-5}\, \mathrm{eV}^2
\leq  \Delta m_{12}^2 \leq  9.1 \times 10^{-5}\, \mathrm{eV}^2\ ,   \quad 
 0.23 \leq \sin^2{\theta_{12}} \leq 0.38\ , \nonumber \\
& &  1.4 \times 10^{-3}\, \mathrm{eV}^2
 \leq  \Delta m_{13}^2  \leq  3.3 \times 10^{-3}\, \mathrm{eV}^2\ ,  \quad
   \sin^2{2 \theta_{23}} \geq 0.90 \ , \nonumber \\
& & \sin^2\theta_{13} \leq  0.047 \ .
\label{data}
\end{eqnarray}

We can also discuss the neutrinoless double beta decay rate, 
which is determined by an effective Majorana mass:
\begin{equation}
\langle m \rangle_{ee}=\left|\ 
m_1c_{12}^2c_{13}^2e^{i \rho}+m_2s_{12}^2c_{13}^2e^{i\sigma}
+m_3s_{13}^2e^{-2i\delta_D} \ \right|\ ,
\end{equation}
\noindent
where $c_{ij}$ and  $s_{ij}$ denote
 $\cos \theta_{ij}$ and $\sin \theta_{ij}$, respectively, 
$\delta_D$ is a so called Dirac phase, and $\rho,\sigma $ are Majorana phases.
Because of $\rho, \sigma\ll 1$ as discussed after eq.(\ref{angle}) 
and $m_1\simeq m_2$, the predicted   $\langle m \rangle_{ee}$ is 
\begin{equation}
\langle m \rangle_{ee}\simeq 
m_1  \simeq  50 \ {\rm meV} \ ,
\end{equation}
which will be tested in future experiments.

\subsection{Right-handed Majorana Neutrino  Mass Matrix}

Let us examine  the right-handed Majorana neutrino mass matrix
to discuss the leptogenesis \cite{FY}.
The mass eigenvalues of $M_R$ in eq.(\ref{rightmatrix}) are obtained as
\begin{equation}
m_{R1}\simeq  
   M_0 (\epsilon_2-2 \epsilon_1^2+\epsilon_3^2-2\epsilon_1^2\epsilon_2) \ ,
              \quad
m_{R2}\simeq    M_0 (\epsilon_2-\epsilon_3^2) \ ,
              \quad
m_{R3}\simeq    M_0 (1+2 \epsilon_1^2+2\epsilon_1^2\epsilon_2) \ ,
\label{rightmass}
\end{equation}
\noindent where the right-handed Majorana neutrinos of 
the first and second family are almost degenerated.
The  mass difference of $m_{R1}$ and  $m_{R2}$ is given by
\begin{equation}
m_{R2}-m_{R1}\simeq   2  M_0 [\epsilon_1^2(1+\epsilon_2) -\epsilon_3^2] \ ,
\label{massdiff}
\end{equation}
which is expected to be 
${\mathcal O}(M_0\epsilon^2\sim M_0\epsilon^3)$ depending on the ratio 
$\epsilon_1/\epsilon_3$.

The orthogonal matrix $O_R$, which diagonalizes the  matrix  $M_R$,  
is given by
\begin{equation}
O_R \simeq 
\left (\matrix{\frac{1}{\sqrt{2}}+\epsilon_x&\frac{1}{\sqrt{2}}&-\epsilon_1\cr
-\left (\frac{1}{\sqrt{2}}+\epsilon_x\right )&\frac{1}{\sqrt{2}}&\epsilon_1\cr
                  \epsilon_1  & 0  & 1\cr
                                         } \right )\ ,
\qquad {\rm with}\qquad
O_R^T M_R O_R = M_R^{\rm diagonal} \ ,
\end{equation}
\noindent
where $\epsilon_x$ is expressed   in terms of  $\epsilon_i$.

 In order to discuss the leptogenesis in the next section,
we take the diagonal basis of the right-handed Majorana neutrino mass matrix.
Then, the Dirac neutrino mass matrix $h_\nu$ is converted  to $O^T_R h_\nu$.
However, for our convenience, we use  $\overline h_\nu= O^T_R h_\nu O_R$
instead of $O^T_R h_\nu$:
\begin{equation}
\overline h_\nu \simeq  h_0
\left (\matrix{1+\frac{\delta_1}{2}-|\epsilon|e^{i\varphi}+2\sqrt{2}\epsilon_x&
 -\frac{\delta_1}{2}&(1-\sqrt{2})\epsilon_1\cr
-\frac{\delta_1}{2} &1+\frac{\delta_1}{2}+|\epsilon| e^{i\varphi} &
 \sqrt{2} |\epsilon| e^{i\varphi}\cr
        (1-\sqrt{2})\epsilon_1& \sqrt{2}|\epsilon| e^{i\varphi}&1+\delta_2\cr
                                         } \right )\ .
\end{equation}
\noindent
The product of $\overline h_\nu\overline h_\nu^\dagger$ appears  
in  the calculations of the leptogenesis.

\section{Leptogenesis}

We consider the leptogenesis \cite{FY,gravitino} via decays of the right-handed 
neutrinos
$N_{Ri}$ which are produced non-thermally by decays of the inflaton
$\varphi_{inf}$ ~\cite{lep-inf}. 

If the phase $\varphi$ does not  vanish, CP invariance is vaiolated 
in the Yukawa matrix $h_\nu$. Then, the interference
between decay amplitudes of tree and one-loop diagrams results in the 
lepton number production \cite{FY}.
 The lepton number asymmetry per decay of the right-handed neutrino $N_{Ri}$
is given by \cite{FY,epsilon1}
\begin{eqnarray}
 \epsilon_{\ell i}
  &\equiv&
  \frac
  {\sum_j\Gamma (N_{Ri}\to l_{Lj} + H) 
  - 
  \sum_j\Gamma (N_{Ri}\to \overline{l_{Lj}} + \overline{H})}
  {\sum_j\Gamma (N_{Ri}\to l_{Lj} + H) 
  +
  \sum_j\Gamma (N_{Ri}\to \overline{l_{Lj}} + \overline{H})}
  \nonumber\\
 &=&
  -\frac{1}{8\pi}
  \frac{1}{( \overline h_\nu  \overline h_\nu^{\dagger})_{ii}}
  \sum_{k\ne i}
  {\rm Im}
  \left[
   \{
   \left(
    \overline h_\nu  \overline h_\nu^{\dagger}
    \right)_{ik}
    \}^2
   \right]
   \left[
    {\cal{F}}_{V}\left(\frac{M_k^2}{M_i^2}\right)
    +
    {\cal{F}}_{S}\left(\frac{M_k^2}{M_i^2}\right)
    \right]\,,
    \label{EQ-epsilon_i}
\end{eqnarray}
where $N_{Ri}$, $l_{Lj}$, and $H$ ($\overline{l_{Lj}}$ and
$\overline{H}$) symbolically denote fermionic or scalar components of
corresponding supermultiplets (and their anti-particles), and
${\cal{F}}_{V}(x)$ and ${\cal{F}}_{S}(x)$ represent  contributions
from vertex and self-energy diagrams, respectively.

In the case of the
SUSY theory, they are given by~\cite{epsilon1-SUSY}
\begin{eqnarray}
 {\cal{F}}_{V}(x) = \sqrt{x}\ln\left( 1 + \frac{1}{x}\right)
  \,,
  \qquad
  {\cal{F}}_{S}(x) = \frac{2\sqrt{x}}{x - 1}
  \, .
  \label{EQ-fs}
\end{eqnarray}
\noindent
Here, we have assumed that the mass difference of the right-handed
neutrinos is large enough compared with their decay widths so that the
perturbative calculation is ensured.

The relevant Yukawa  couplings are given as 
\begin{eqnarray}
&&  {\rm Im}[\{(\overline h_\nu  \overline h_\nu^\dagger)_{12}\}^2]= 
-{\rm Im}[\{(\overline h_\nu  \overline h_\nu^\dagger)_{21}\}^2]=
  -2 \  |h_0|^4 \ |\epsilon| \ \delta_1^2 \ 
\left [1+(2-\sqrt{2})\frac{\epsilon_1}{\delta_1} \right ]\sin\varphi \ ,
\nonumber \\
&&  {\rm Im}[\{(\overline h_\nu  \overline h_\nu^\dagger)_{13}\}^2]= 
  2 \  |h_0|^4 \ |\epsilon| \ \epsilon_1 \ 
\left [2(2\sqrt{2}-3)\epsilon_1-(2-\sqrt{2})\delta_1 \right ]\sin\varphi \ ,
\nonumber \\
&&  {\rm Im}[\{(\overline h_\nu  \overline h_\nu^\dagger)_{23}\}^2]= 
  2 \  |h_0|^4 \ |\epsilon|^2  \ (2\delta_2-\delta_1 )\sin 2\varphi \ ,
\nonumber \\
&&(\overline h_\nu  \overline h_\nu^\dagger)_{11}
= |h_0|^2 (1+\delta_1-2|\epsilon| \cos\varphi) \ ,
\nonumber\\
&&(\overline  h_\nu  \overline h_\nu^\dagger)_{22}= 
|h_0|^2 (1+\delta_1+2|\epsilon| \cos\varphi)\ .
\label{hhdagger}
\end{eqnarray}
Since the right-handed neutrino masses of 
the first and second families  are almost degenerate  in the present model
\footnote{The leptogenesis with almost degenerate Majorana neutrinos 
has been examined in the framework of the democratic mass matrix 
\cite{Almost}.},
the self-energy contribution  ${\cal{F}}_{S}(x)$ is much larger than the 
vertex contribution $ {\cal{F}}_{V}(x)$ as follows:
\begin{eqnarray}
&& {\cal{F}}_{V}(\frac{m_{R2}^2}{m_{R1}^2})\simeq
 {\cal{F}}_{V}(\frac{m_{R1}^2}{m_{R2}^2})\simeq \ln 2 \ ,
\nonumber \\
&& {\cal{F}}_{S}(\frac{m_{R2}^2}{m_{R1}^2})= 
-{\cal{F}}_{S}(\frac{m_{R1}^2}{m_{R2}^2})= 
      \frac{2m_{R1}m_{R2}}{m_{R2}^2-m_{R1}^2}\simeq \frac{1}{2}
\frac{\epsilon_2(\epsilon_2-2\epsilon_1^2)}
{\epsilon_2(\epsilon_1^2-\epsilon_3^2)+\epsilon_1^2(\epsilon_2^2+
\epsilon_3^2-\epsilon_1^2)}\ .
\end{eqnarray}
\noindent
Therefore,  we have 
\begin{eqnarray}
{\cal{F}}_{S}(\frac{m_{R2}^2}{m_{R1}^2})= 
-{\cal{F}}_{S}(\frac{m_{R1}^2}{m_{R2}^2})
      = {\mathcal O}(\epsilon_2^{-1}\sim\epsilon_2^{-2}) \ ,
\end{eqnarray}
 depending on the ratio $\epsilon_1/\epsilon_3$.
Notice that a large enhancement is obtained in the case of
  $\epsilon_1=\epsilon_3$.

The lepton number asymmetry is given
\begin{eqnarray}
 \epsilon_{\ell 1}= \epsilon_{\ell 2}\simeq 
\frac{|h_0|^2}{4\pi} |\epsilon|\delta_1^2
\left (2-\sqrt{2}\frac{\epsilon_1}{\delta_1}\right )
\frac{\epsilon_2(\epsilon_2-2\epsilon_1^2)}
{\epsilon_2(\epsilon_1^2-\epsilon_3^2)+\epsilon_1^2(\epsilon_2^2+
\epsilon_3^2-\epsilon_1^2)}\sin\varphi \ ,
\label{asym}
\end{eqnarray}
which turns out 
\begin{eqnarray}
 \epsilon_{\ell 1}
= \epsilon_{\ell 2}\simeq \frac{|h_0|^2}{2\pi} \delta_1^2\sin\varphi \ .
\end{eqnarray}
\noindent
The value of $|h_0|^2$ is given in terms of $m_1$ and $m_{R1}$:
\begin{eqnarray}
 |h_0|^2
\simeq  \frac{m_{1} m_{R1}}{(174\ \sin \beta \ {\rm GeV})^2}
\simeq \frac{1.65}{\sin^2 \beta} \times 10^{-15} \frac{m_{R1}}{1 {\rm GeV}} \ ,
\end{eqnarray}
\noindent
where  the $vev$ of the Higgs $H_u$, $174\sin\beta  {\rm GeV} $, 
and  $m_1=0.05 {\rm eV}$ are used.
The ratio of the lepton number density $n_L$ to the entropy density $s$
produced by the inflaton decay is given by~\cite{lep-inf}
\begin{eqnarray}
 \frac{n_L}{s} = \frac{3}{2}\sum_i
  \epsilon_{\ell i}
  B_r^{(i)}
  \frac{T_R}{m_{\phi_{inf}}}
  \,,
\end{eqnarray}
where $T_R$ is the reheating temperature after the inflation, $m_{\phi_{inf}}$
the mass of the inflaton, and $B_r^{(i)}$ the branching ratio of the
decay channel of the inflaton to $N_{Ri}$, i.e., $B_r^{(i)} =
B_r(\phi\to N_{Ri}N_{Ri})$. Here, we have assumed that the inflaton
decays into a pair of right-handed neutrinos, and $M_{Ri}>T_R$ in order
to make the generated lepton asymmetry not washed out by lepton-number
violating processes after the $N_{Ri}$'s decay. 
A part of the produced lepton asymmetry is immediately
converted~\cite{FY} into the baryon asymmetry via the ``sphaleron''
effect~\cite{sphaleron}, since the decays of $N_{Ri}$ take place much
before the electroweak phase transition. The baryon asymmetry is given by
\begin{eqnarray}
 \frac{n_B}{s} = C\frac{n_L}{s}
  \,,
\end{eqnarray}
where $C$ is given by $C \simeq - 0.35$ in the minimal SUSY standard
model ~\cite{LtoB}.

Therefore, the amount of the baryon asymmetry in the present model is
estimated as
\begin{eqnarray}
 \frac{n_B}{s}
  &\simeq&
 - \frac{0.7}{\sin^2 \beta}\times 10^{-6}
  \  \delta_1^2 \sin\varphi
  \left( B_r^{(2)} + B_r^{(3)} \right)
   \left(\frac{T_R}{10^{10}{\rm GeV}}\right)
  \left(\frac{2 {\cal{M}}_R}{m_{\phi_{inf}}}\right) \ .
  \label{EQ-nBs}
\end{eqnarray}
We see that the observed baryon asymmetry $n_B/s \simeq (0.8$--$0.9)\times
10^{-10}$~\cite{PDG} is obtained for a  choice of the parameter,
$T_R=10^{10}{\rm GeV}$ and $\delta_1^2=10^{-4}$ with $\varphi=-\pi/4$. 

We should note  a possibility of  enhancement of $\epsilon_{\ell 1}$
and  $\epsilon_{\ell 2}$.
The lepton number asymmetries  in the eq.(\ref{asym}) are  enhanced
in the case of  $\epsilon_1=\epsilon_3$, where we obtain
\begin{eqnarray}
 \epsilon_{\ell 1}
= \epsilon_{\ell 2}\simeq \frac{|h_0|^2}{2\pi} \frac{\delta_1^2}{|\epsilon|}
\sin\varphi \ ,
\end{eqnarray}
where $\epsilon_2=\epsilon_1=|\epsilon|$ is taken.
Therefore, the baryon asymmetry $n_B/s$ is enhanced by the order of  $10^3$
because of  $|\epsilon|\simeq 10^{-3}$.
Then,   $T_R=10^{7}{\rm GeV}$  is allowed to reproduce 
the observed baryon asymmetry.

\section{Summary}

We have discussed the neutrino mass matrix in a (1+5) dimensional space-time
where  two extra dimensions are compactified on a ${\bf T^2/Z_3}$ orbifold 
\cite{Geometry, IIB}. Here, a ${\bf 5}^*$ and a right-handed neutrino $N$  
in each family are localized on each of the equivalent three fixed points 
of the ${\bf T^2/Z_3}$ orbifold while three ${\bf 10}$'s and Higgs doublets 
$H_u$ and $H_d$ live in the bulk. We have considered, in this paper, 
that the Higgs field $\phi$ responsible for the B-L breaking is localized on 
the fixed point of the third family of ${\bf 5}^*$ and $N$.
This setting leads to a hierarchy in the Majorana mass spectrum for the right-handed neutrinos, $m_{R1}\simeq m_{R2}\ll m_{R3}$,
 which naturally generates
  the inverted  hierarchy of the neutrino masses.

 
The  present model  explains the deviation of the $\theta_{12}$
from the maximal mixing. It also predicts the element
of the neutrino mass matrix, $\langle m\rangle_{ee}$, responsible for 
neutrinoless double beta decays as $\langle m\rangle_{ee}\simeq 50$ meV,
which   will be accessible to future experiments.
The mixing angle $\theta_{13}$  
strongly depends on $S_3$ breaking terms. 
 
We have shown that 
the observed  baryon asymmetry in the present universe is
produced by the non-thermal leptogenesis via the inflaton decay.
Due to $m_{R1}\simeq m_{R2}$ the baryon asymmetry is enhanced. 
In conclusion, we have found that the leptogenesis  works well 
 with the reheating temperature  $T_R=10^{7}\sim 10^{10} {\rm GeV}$.

\paragraph{Acknowledgement}

The work of M.T.\ has been  supported by the
Grant-in-Aid for Science Research
of the Ministry of Education, Science, and Culture of Japan
No.\ 16028205, No.\ 17540243.
The work of T.Y.\ has been supported in part by a Humboldt Research Award.


\end{document}